\documentclass[reprint,
 amsmath,amssymb,
 aps,
prstab
]{revtex4-1}

\usepackage{graphicx}
\usepackage{subcaption}
\usepackage{stackengine}
\usepackage{dcolumn}
\usepackage{bm}
\usepackage{xcolor}
\usepackage{comment}

\usepackage{amsmath}

\usepackage{textcomp} 

\makeatletter
\newcommand\thefontsize[1]{{#1 The current font size is: \f@size pt\par}}
\newcommand\thefontsizeHere{{The current font size is: \f@size pt\par}}
\makeatother

\begin{document}

\preprint{}

\title{Fast Tracking of 6D Particle Phase Space Using the FC2CT Algorithm for rf Cavities}

\author{M.~Southerby}
\email{m.southerby@lancaster.ac.uk}
\author{R.~Apsimon}
\email{r.apsimon@lancaster.ac.uk}
\affiliation{Engineering Department, Lancaster University, Lancaster, LA1 4YW, UK }
\affiliation{Cockcroft Institute, Daresbury Laboratory, Warrington, WA4 4AD, UK}

\date{\today}

\begin{abstract}

In this paper, a fast cell to cell tracking algorithm (FC2CT) is developed, that determines the change in 6D phase space of a particle beam through rf accelerating cavities operated in both Standing and Traveling Wave modes. The performance is compared to well trusted tracking codes - ASTRA and RF-Track - for proton beams with relativistic beta $\approx$ 0.5. The FC2CT algorithm produces a complete analytical method to determine the change in particle momentum, with the constant particle velocity approximation over the integration region. Additional developments for FC2CT are discussed, such as altering the integration length. FC2CT can be used as a fast method class in common tracking codes to perform fast simulations through rf cavities.
\end{abstract}

\maketitle

\section{Introduction}

The use of particle accelerators are paramount in multiple industries, examples including the health care, cargo scanning and research industries. As a result, there is constant requirement for additional developments within the particle accelerator industry, ensuring optimal efficiency of machines or increased beam energies. A large subset of accelerators are linear accelerators (linacs) that accelerate particles using radio-frequency (rf) cavities that contain electromagnetic (em) fields for particle acceleration. Beam dynamics simulations are a vital aspect of accelerator design, however can be computationally expensive, for example beam dynamics studies for circular machines requiring many thousands of turns.

In this paper, we outline a fast tracking algorithm for beam dynamics through coupled cavity linacs (CCL), and is called the Fast Cell to Cell Tracking algorithm (FC2CT), calculating the change in 6D phase space of a particle over one rf cell per iteration. The model approximates the particle velocity is constant over an rf cell, and space charge effects are ignored. The first approximation is accurate for protons of all energy that are to be accelerated by a CCL. The proton transition energy for more efficient acceleration by CCL is often taken to be $\approx$~80-100~MeV, below which drift tube linacs may have increased shunt impedance \cite{shunt_impedance_diff_strucs}. Nevertheless, some CCL accelerate protons from as low as 37.5 MeV \cite{light_linac}. FC2CT is only accurate for electrons over $\approx$ 3 MeV. The tracking algorithm requires the on axis longitudinal electric field, from which the remaining non-zero field components ($E_r, B_{\theta}$) are approximated by determining the Taylor expansion coefficients in the small radial limit. This method has been utilised before, in order to quickly provide a method for 3D particle tracking in well-known tracking codes, such as ASTRA \cite{astra}. The method is accurate for both Standing Wave (SW) and Traveling wave (TW) cavity types, however require slightly different algorithms. In addition, the algorithm described is purely for the TM$_{010}$ cavity mode. The accuracy of the FC2CT model is compared with well trusted tracking codes, ASTRA (SW) and RF-Track (TW), \cite{rftrack}.

\subsection{Lorentz Force law and Maxwell Equations for a TM$_{010}$ Mode}

A particle in a region of non-zero em field will experience a force given by the Lorentz force \cite{griffiths}. The force is the rate of change of momentum;
\begin{equation}
    \Delta \vec{P} = \int d\vec{P} = q \int \left(\vec{E}(t) + \vec{v}(t) \times \vec{B}(t)\right) dt.
    \label{eq:lorentz}
\end{equation}
Where $\vec{v}(t)$ is the particle velocity. In general, the em field and particle velocity are a function of time. The em fields in a given region will be described by Maxwell equations \cite{griffiths}. For an rf single cell, the TM$_{010}$ mode is often used for acceleration \cite{wangler}. The longitudinal electric field ($E_z$) and azimuthal magnetic field ($B_{\theta}$) in a TM$_{010}$ mode for a simple pillbox cavity are described by the zeroth and first order Bessel functions of the first kind \cite{bessel}. rf single cells are often designed with nose-cones, to concentrate $E_z$ on-axis and improve shunt impedance and require a beam pipe, to allow power coupling and particle transmission. As a result, the initially separable em field components in the case of a pillbox cavity become functions of the longitudinal displacement, $z$.

As a particle traversing an rf single cell will be at a small radial displacement relative to the cell radius, it is possible to describe the non-zero em field components as a function of the on-axis $E_z$ component, denoted~$E_z(r=0, t)$, using a Taylor expansion about $r$ = 0. The TM$_{010}$ mode in a given rf cell will have negligible $E_{\theta}$, $B_{r}$, $B_{z}$ components, as in a pillbox cavity.

The non-zero em field components observed by a particle at position, $z$, time, $t$ can be described as follows,
\begin{multline}
    E_z(r, z, t) = \\ \left[E_z(0, z) -\frac{r^2}{4}\left(\frac{\partial ^2 E_z}{\partial z ^2} +  \frac{\omega^2}{c^2} E_z(0, z) \right) \right]\textrm{cos}\left(\phi_0 +\omega t\right),
    \label{eq:Ez_taylor}
\end{multline}

\begin{multline}
    E_r(r, z, t) = \\ \left[ -\frac{r}{2} \frac{\partial E_z}{\partial z} + \frac{r^3}{16} \left( \frac{\partial ^3 E_z}{\partial z^3} + \frac{\omega^2}{c^2} \frac{\partial E_z}{\partial z} \right) \right]\textrm{cos}\left(\phi_0 + \omega t\right),
    \label{eq:Er_taylor}
\end{multline}

\begin{multline}
    B_{\theta}(r, z, t) = \\ \frac{\omega}{c^2}\left[ \frac{r}{2} E_z(0, z) - \frac{r^3}{16} \left( \frac{\partial ^2 E_z}{\partial z^2} + \frac{\omega ^2 }{c^2} E_z(0, z)\right)  \right] \\ \times \textrm{cos}\left(\phi_{0} - \frac{\pi}{2} + \omega t\right).
    \label{eq:Bphi_taylor}
\end{multline}

Where $\phi_0$ is the rf phase when the particle enters the cell (at $z=0$). The magnetic field lags the electric field by $\pi/2$. $E_z(0, z)$ is the longitudinal on-axis electric field component, and can be exported from em solver codes, such as CST microwave solver \cite{CST} for a given rf single cell geometry. The terms $\frac{\partial^n E_z}{\partial z^n}$ describe the $n_{th}$ derivative of $E_z(0, z)$.

\begin{figure}[h!]
\centering
     \subcaptionbox{}{\includegraphics[width=2.5in]{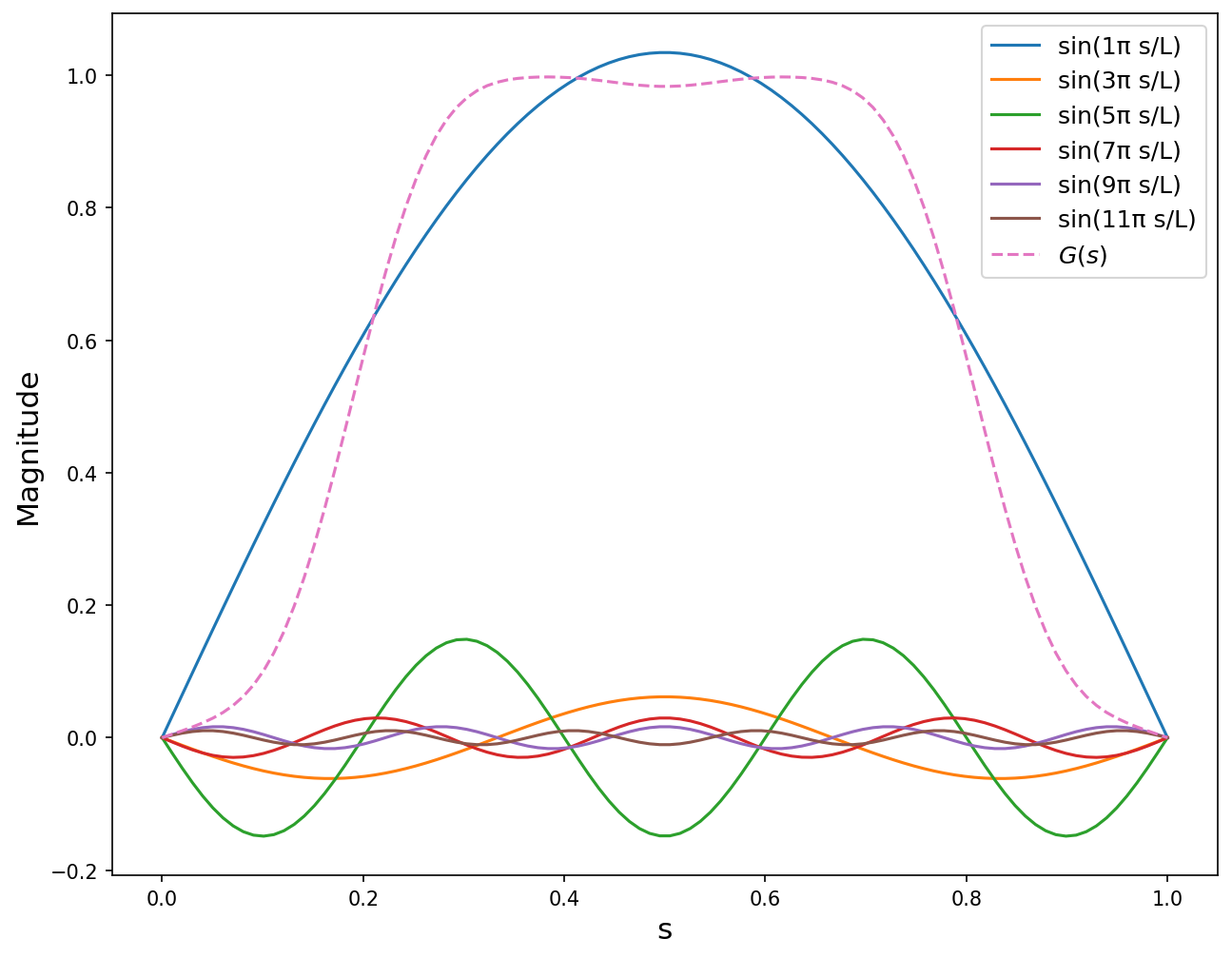}}
     \caption{Normalised on-axis $E_z$ component, G(s) and the individual Fourier components.}
     \label{fig:Fourier_field}
\end{figure}

It is required to describe the on-axis $E_z$ component as a Fourier series \cite{fourier}, as $E_z$ is periodic over some period, $P$, defined by the phase advance per cell of the cavity. As $E_z(0, z)$ can be described by a Fourier series, the partial derivatives of $E_z$ become straight-forward to compute. The Fourier terms required to construct $E_z$ for a SW cell is shown in Fig.~\ref{fig:Fourier_field}.

For a SW cavity, the longitudinal electric field, and the derivatives, are given below at time $t$;
\begin{equation}
    E_z(r=0, z, t) = \sum_{n=1}^{N} b_n \hspace{1mm}\textrm{sin}\left(\frac{2n \pi z}{P}\right)\cos\left( \omega t + \phi_0\right),,
    \label{eq:sw_field}
\end{equation}
\begin{equation}
    \frac{\partial E_z}{\partial z} = \left(\frac{2\pi}{P}\right)\sum_{n=1}^{N} n b_n \hspace{1mm} \textrm{cos}\left(\frac{2n \pi z}{P}\right)\cos\left(\omega t + \phi_0\right)
    \label{eq:sw_field_deriv}
\end{equation}
\begin{equation}
    \frac{\partial ^2 E_z}{\partial z^2} = -\left(\frac{2\pi}{P}\right)^2\sum_{n=1}^{N} n^2 b_n \hspace{1mm} \textrm{sin}\left(\frac{2n \pi z}{P}\right)\cos\left(\omega t + \phi_0\right).
    \label{eq:sw_field_deriv2}
\end{equation}
Where $b_n$ are the Fourier series coefficients. The required number of coefficients varies for different field profiles, and should be the minimum number such that the field is accurately represented.

When the operating mode is a $\frac{\pi}{2}$-mode, the field period is two cell lengths, $P = 2 L_{cell}$. Thus, $E_z$ can be described using only $b_n$ terms, as $a_n = 0 $ for all $n$. For TW cavities, the longitudinal electric field has both real ($A(z)$) and imaginary ($B(z)$) components, and thus both components (over one period) must be imported from an em field solver. Both the real and imaginary fields can also be described as a Fourier series, however will require both $a_n$ and $b_n$ coefficients. The field seen by a particle in a TW cavity is the real component of the complex field, and is given;
\begin{equation}
    E_z(r=0, z, t) = A(z) \cos\left(\omega t + \phi_0 \right) - B(z)  \sin\left( \omega t + \phi_0\right).
    \label{eq:tw_field}
\end{equation}
Where A(z) and B(z) are described by Fourier series;
\begin{equation}
    A(z) = \sum_{n=1}^{N} b_{n}^{\text{real}}\sin\left(\frac{2n \pi z}{P}\right) + a_{n}^{\text{real}} \cos\left(\frac{2n \pi z}{P}\right),
\end{equation}
and
\begin{equation}
    B(z) = \sum_{n=1}^{N} b_{n}^{\text{imag}}\sin\left(\frac{2n \pi z}{P}\right) + a_{n}^{\text{imag}}\cos\left(\frac{2n \pi z}{P}\right).
\end{equation}
From Eqn.~\ref{eq:lorentz}, the momentum change in the $z$ direction is given;
\begin{equation}
    \Delta P_z = q \int (E_z(r, z, t) + v_x B_y(r, z, t) - v_y B_x(r, z, t) ) dt.
    \label{eq:delta_pz}
\end{equation}

As the particle velocity over the rf cell ($v_z$) is approximately constant, $dt = \frac{dz}{v_z}$. It is also assumed that terms including $\frac{v_x}{v_z}$ are negligible, thus;

\begin{equation}
    \Delta P_z = \frac{q}{\beta_z c} \int E_z(r, z, t) dz.
    \label{eq:delta_pz_simplified}
\end{equation}

\begin{figure}[h!]
\centering
     \subcaptionbox{}{\includegraphics[width=2.5in]{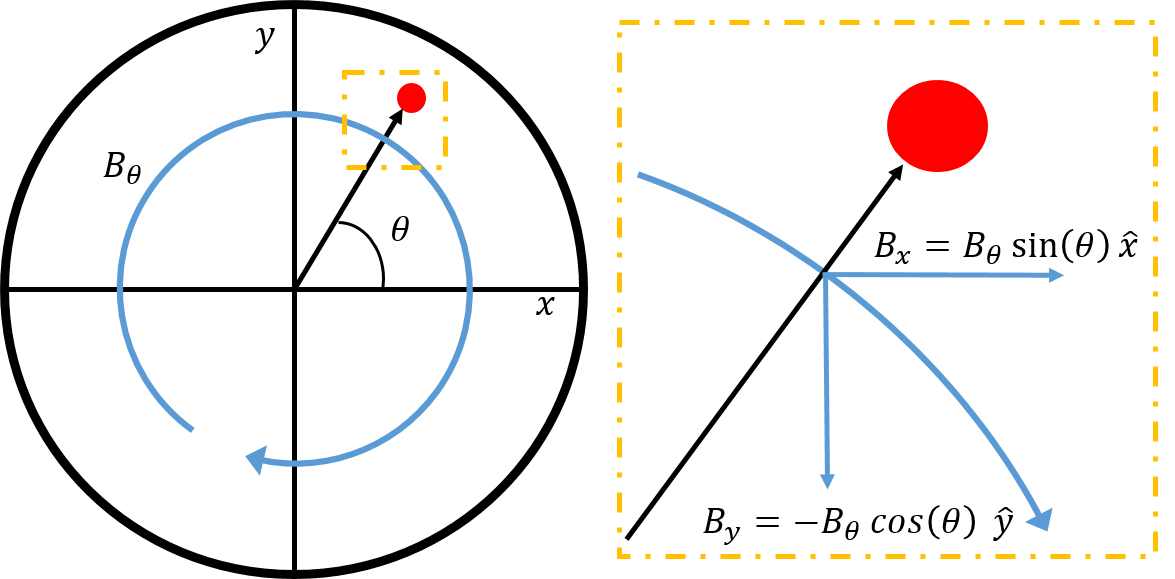}}
     \caption{Schematic showing azimuthal angle, $\theta$.}
     \label{fig:theta}
\end{figure}

The momentum change in the transverse planes are also given by the Lorentz force, and are a function of the particles azimuthal angle, $\theta$, depicted in Fig.~\ref{fig:theta}.

\begin{equation}
    \Delta P_{x}  = qc\cos\left(\theta\right)  \left(\frac{1}{\beta_z c}\int E_r dz + \int B_{\theta} dz \right)
    \label{eq:lorentz_px}
\end{equation}
\begin{equation}
    \Delta P_{y}  = qc\sin\left(\theta\right) \left(\frac{1}{\beta_z c}\int E_r dz + \int B_\theta dz \right) .
    \label{eq:lorentz_py}
\end{equation}

\subsection{Derivation of single cell SW}
\label{sec:sw_gain}

Combining Eqns.~\ref{eq:Ez_taylor} and \ref{eq:delta_pz_simplified} the change in $P_z$ is given;
\begin{multline}
    \Delta P_z = \\  \frac{q}{\beta_z c} \int_{0}^{L_{cell}}\left[E_z(0, z) -\frac{r^2}{4}\left(\frac{\partial ^2 E_z}{\partial z ^2} + \frac{\omega^2}{c^2} E_z(0, z) \right) \right] \\ \times \textrm{cos}\left(\phi_0 + \frac{\omega z}{\beta_z c}\right) dz,
    \label{eq:delta_pz2}
\end{multline}
where the integration is defined over a single rf cell (the integration range can be over multiple cells, or sections of a single rf cell). Equation~\ref{eq:delta_pz2} can be simplified by noting all terms are of the form;
\begin{multline}
    F(b_{n}, i, \beta_s/\beta_z, \phi_0)  =  \\  \int_0^{L_{cell}}\frac{\partial^i}{\partial z^i}[E_z] \cos\left(\phi_0 + \frac{\omega z}{\beta_z c} \right) dz.
    \label{eq:the_trick}
\end{multline}
Therefore, using the standard result of Eqn.~\ref{eq:the_trick}, Eqn.~\ref{eq:delta_pz2} is simplified to
\begin{multline}
    \Delta  P_z = \\  \frac{q}{\beta_z c} \left(1  - \frac{\omega^2}{c^2} \frac{r^2}{4}\right) F(b_{n}, 0, \beta_s/\beta_z, \phi_0) - \\ \frac{q}{\beta_z c} \frac{r^2}{4} F(b_{n}, 2, \beta_s/\beta_z, \phi_0).
    \label{eq:delta_pz3}
\end{multline}
Where
\begin{equation}
    \begin{split}
     F\biggl(b_{n}, {\stackanchor{\emph{i} even}{\emph{i} odd}}, \beta_s/\beta_z, \phi_0\biggr)  = \frac{u(i)}{2} \biggl(\frac{\pi}{L_{cell}} \biggr)^{(i-1)} \times\\ \sum_{n=1}^{\infty}  n^i b_n  \Biggl[\frac{
    {\stackanchor{cos}{sin}}\biggl(\frac{\beta_s}{\beta_z}\pi - n\pi + \phi_0\biggr) - {\stackanchor{cos}{sin}}\bigl(\phi_0\bigr)}{(\frac{\beta_s}{\beta_z} - n)} {\mp} \\ \frac{{\stackanchor{cos}{sin}}\biggl( \frac{\beta_s}{\beta_z}\pi + n\pi + \phi_0\biggr) - {\stackanchor{cos}{sin}}\bigl(\phi_0\bigr)}{(\frac{\beta_s}{\beta_z} + n)}\Biggr]
    \label{eq:F_function}
    \end{split}
\end{equation}

and
\begin{equation}
    u(i)=
    \begin{cases}
      +1, & \text{if}\ i= 0, 1 \\
      -1, & \text{if}\ i = 2, 3.
    \end{cases}
  \end{equation}
Calculating similar forms for Eqns.~\ref{eq:lorentz_px} and \ref{eq:lorentz_py} can be completed via substitution of $E_r$ and $B_{\theta}$ as functions of purely $E_z$ and its derivatives, using Eqns.~\ref{eq:Er_taylor} and \ref{eq:Bphi_taylor}. Finally, the standard result of Eqn.~\ref{eq:the_trick} can be used. For the integral of $B_{\theta}$, the input rf phase, $\phi_0$ becomes $\phi_0 - \frac{\pi}{2}$.

\subsection{Derivation of single cell TW}

The derivation of the momentum gain for TW cavities is more involved, as both the real and imaginary field maps of $E_z(r=0)$ are required. This is because the field behaves as described in Eqn.~\ref{eq:tw_field}. The longitudinal momentum change, $\Delta P_z$, is therefore found by computing the following integral;
\begin{multline}
    \Delta P_z = \\ \frac{q}{\beta_z c} \int \Bigl[A(z) \cos\left(\frac{\omega z}{\beta_z c} + \phi_0 \right) - B(z)  \sin\left( \frac{\omega z}{\beta_z c} + \phi_0\right)\Bigr] - \\ \frac{r^2}{4}\Biggl(\Bigl[\frac{\partial ^2A(z)}{\partial z ^2} \cos\left(\frac{\omega z}{\beta_z c} + \phi_0 \right) - \frac{\partial  ^2 B(z)}{\partial z ^2}  \sin\left( \frac{\omega z}{\beta_z c} + \phi_0\right)\Bigr] + \\  \frac{\omega^2}{c^2} \Bigl[A(z) \cos\left(\frac{\omega z}{\beta_z c} + \phi_0 \right) - B(z)  \sin\left( \frac{\omega z}{\beta_z c} + \phi_0\right)\Bigr] \Biggr) dz.
    \label{eq:delta_pz_TW}
\end{multline}
Using Eqn.~\ref{eq:the_trick}, $\Delta P_z$ can be calculated for the TW case, noting that $\sin(x) = \cos(x - \frac{\pi}{2})$. In general, an iteration over a TW single cell requires four times the number of calculations than a SW cell, due to requiring both $a_n$ and $b_n$ coefficients for describing both the real and imaginary TW field components, $A(z)$ and $B(z)$. As a result, the FC2CT method is less effective for TW cavities, as the speed benefit is reduced.

As before, functional forms for $\Delta P_x$ and $\Delta P_y$ can be determined using Eqns.~\ref{eq:Er_taylor},~\ref{eq:Bphi_taylor},~\ref{eq:the_trick}~and~\ref{eq:F_function}.

\subsection{Calculation of Position in SW and TW cavities}
The 3D particle position is also updated every rf cell. Firstly, the time of flight, $t_{cell}$, is calculated as follows;
\begin{equation}
    t_{cell} = \overline{\beta_z} L_{cell},
\end{equation}
where the average particle velocity along the $i_{th}$ axis is defined
\begin{equation}
    \overline{\beta_i} = \frac{\overline{P_i}}{\sqrt{m^2 + \overline{P_x^2} + \overline{P_y^2}+ \overline{P_z^2}}}.
\end{equation}
The updated positions are therefore;
\begin{equation}
    z = z + L_{cell}, \hspace{2mm} x = x+{t_{cell}} c \overline{\beta_x}, \hspace{2mm} y = y + {t_{cell}} c \overline{\beta_y}.
\end{equation}

For both the SW and TW methods, the rf phase, $\phi_0$, is updated every cell due to phase slippage. The phase slippage is a function of the difference between particle and design velocity;
\begin{equation}
    \Delta \phi = \omega \Delta t 
     \approx \frac{\omega L_{cell}}{\beta_s c} \left(1 - \frac{\beta_s}{\overline{\beta_z}} \right).
    \label{eq:slippage}
\end{equation}

A script was written in Python \cite{python}, that takes the on-axis $E_z$ field map as an input, and extracts the Fourier coefficients. The script tracks a particle beam through an rf cavity with a specified number of the cells, with a specified gradient. The 6D phase space of each particle is updated after every cell, as is the rf phase. The position of the particle is increased by $L_{cell}$ in the longitudinal direction and determined transversely using a calculated time of flight and transverse velocity. This allows a single iteration to be performed each rf cell, as opposed to multiple iterations, as used in conventional tracking methods. Section~\ref{sec:Performance} describes the accuracy of the FC2CT method for both SW for protons with kinetic energy 37.5 and 150 MeV, and TW structures for protons at 150 MeV.

\section{Performance}
\label{sec:Performance}

For the comparison of SW FC2CT, ASTRA tracking code was used. The input distribution was created using the ASTRA generator function, producing protons at 37.5 and 150~MeV. Protons at 37.5~MeV are simulated as below this energy region, proton acceleration is often accomplished with RFQ and drift tube linac \cite{wangler}, as the effective shunt impedance is higher than for CCL. Thus, exploring both 37.5 and 150~MeV proton energies assesses FC2CT for different values of relativistic Lorentz factor, $\gamma_r$. In addition, showing FC2CT is accurate for proton energies of 150~MeV, demonstrates the method will be accurate for higher energy protons (as the only approximation is constant relativistic $\beta$ over a cell).

\begin{figure}[ht!]
\centering
     \subcaptionbox{}{\includegraphics[width=2.5in]{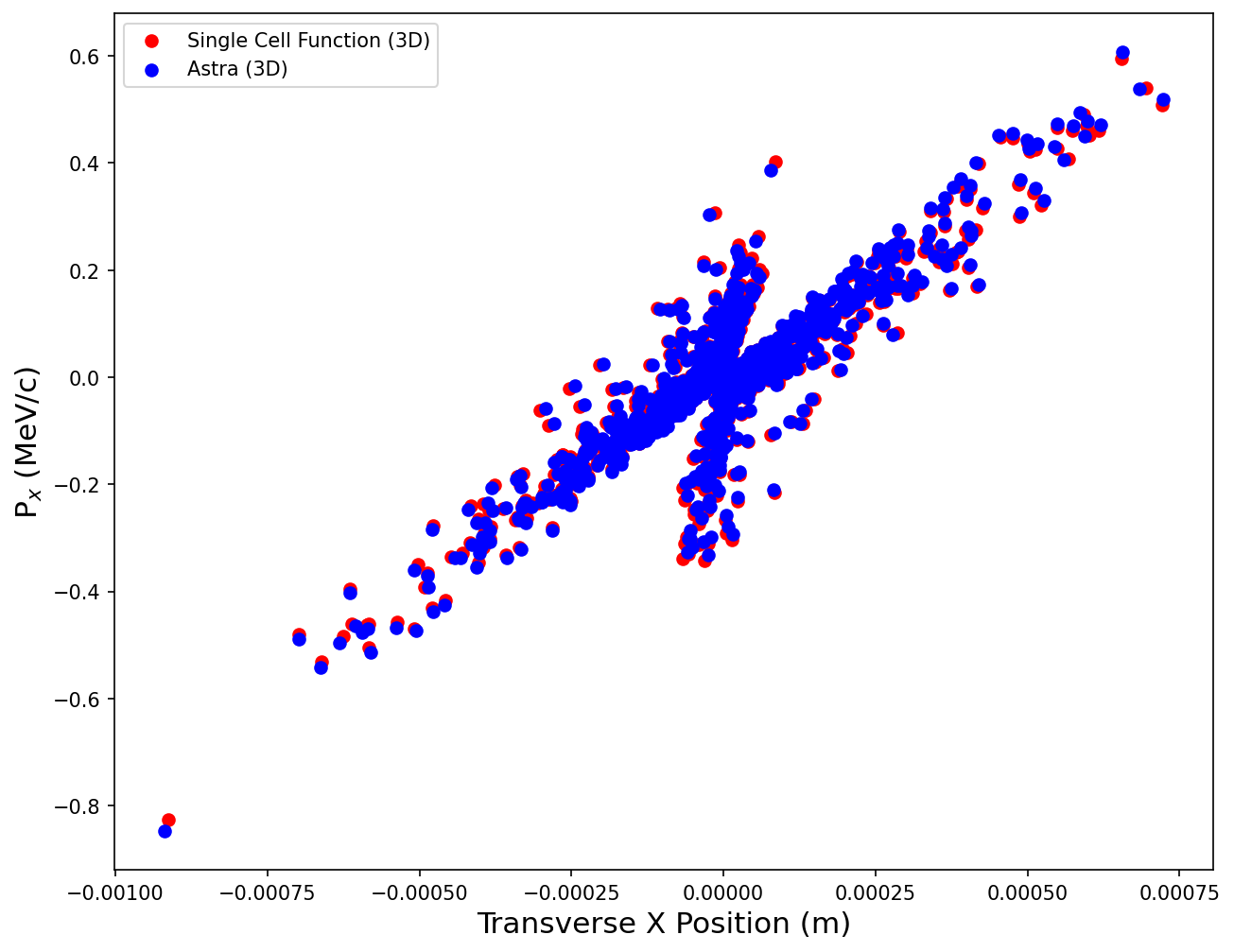}}
     \subcaptionbox{}{\includegraphics[width=2.5in]{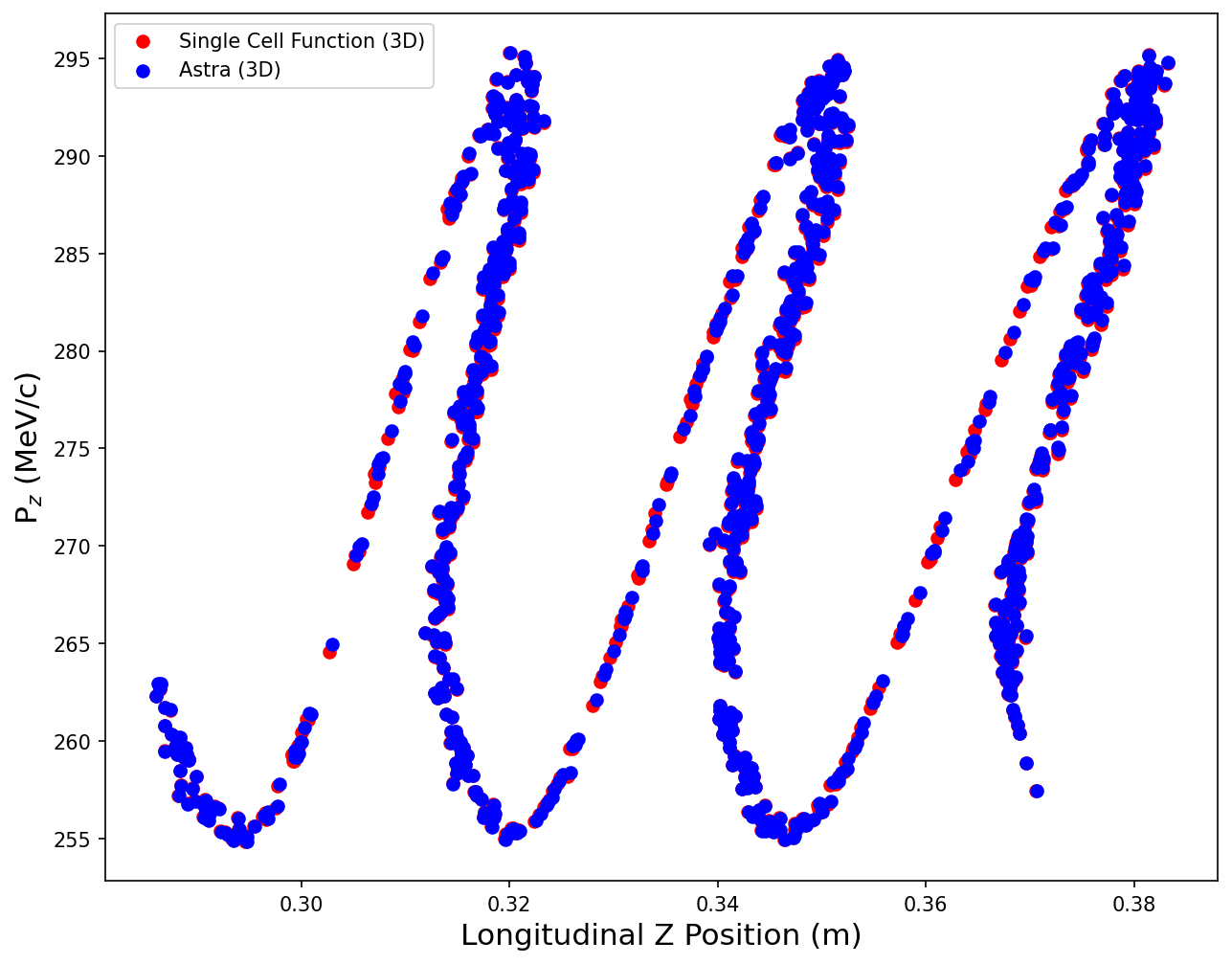}}
     \caption{Transverse and longitudinal phase space of proton beam (initial energy 37.5 MeV) after traversing 20 SW cells, as calculated by ASTRA and FC2CT.}
     \label{fig:37.5_protons}
\end{figure}

Side-coupled single cells were designed in CST Microwave Solver \cite{CST}. The cell lengths were defined such as to keep the field synchronous with protons of energy of 37.5 and 150~MeV. Figure~\ref{fig:37.5_protons} shows the transverse and longitudinal phase space of a proton beam after 20 SW cells. The tracked phase space is shown as calculated by both ASTRA and FC2CT. The two distributions agree strongly, suggesting the constant $\beta$ approximation is valid for protons at 37.5~MeV.

\begin{figure}[ht]
\centering
     \subcaptionbox{}{\includegraphics[width=2.5in]{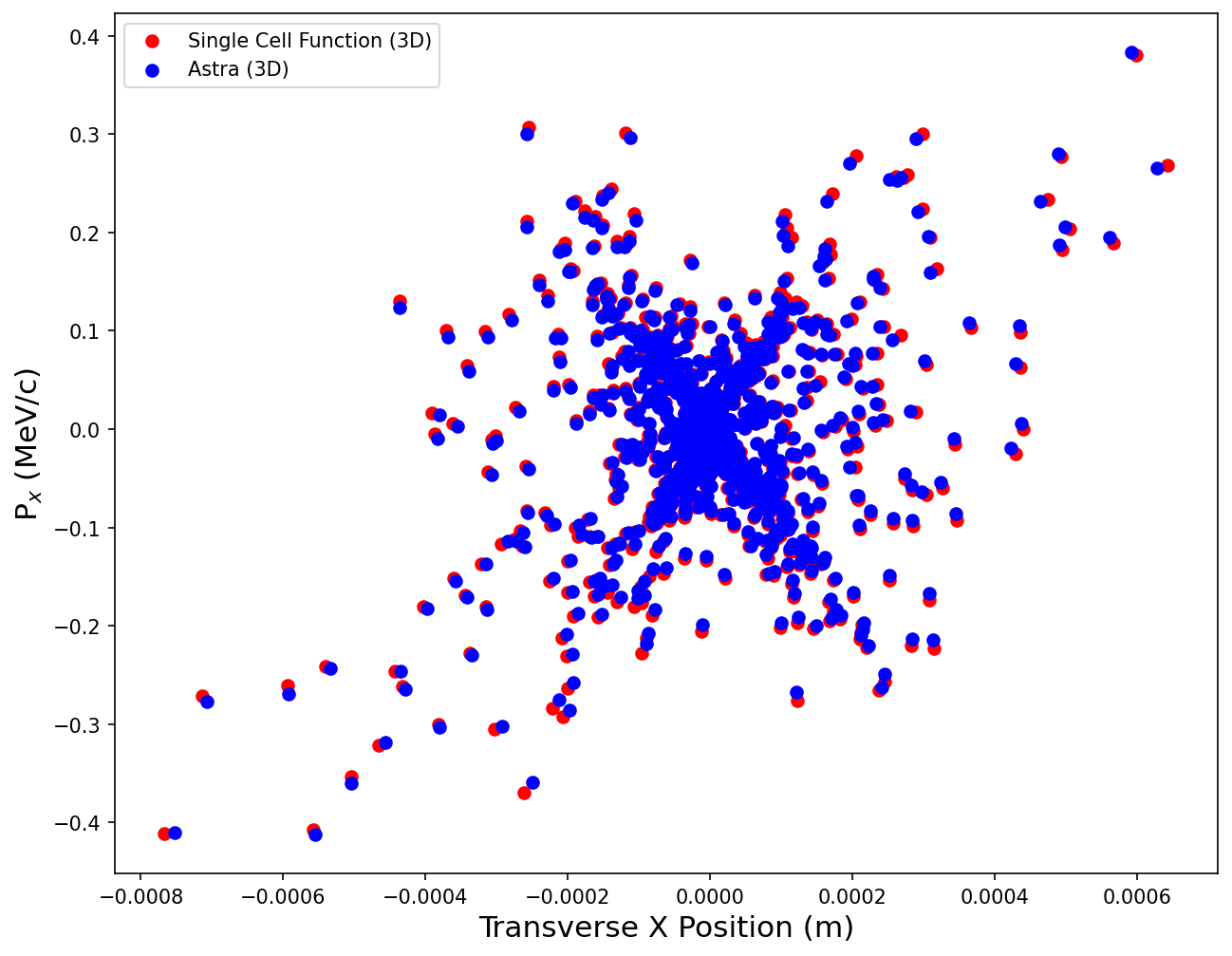}}
     \subcaptionbox{}{\includegraphics[width=2.5in]{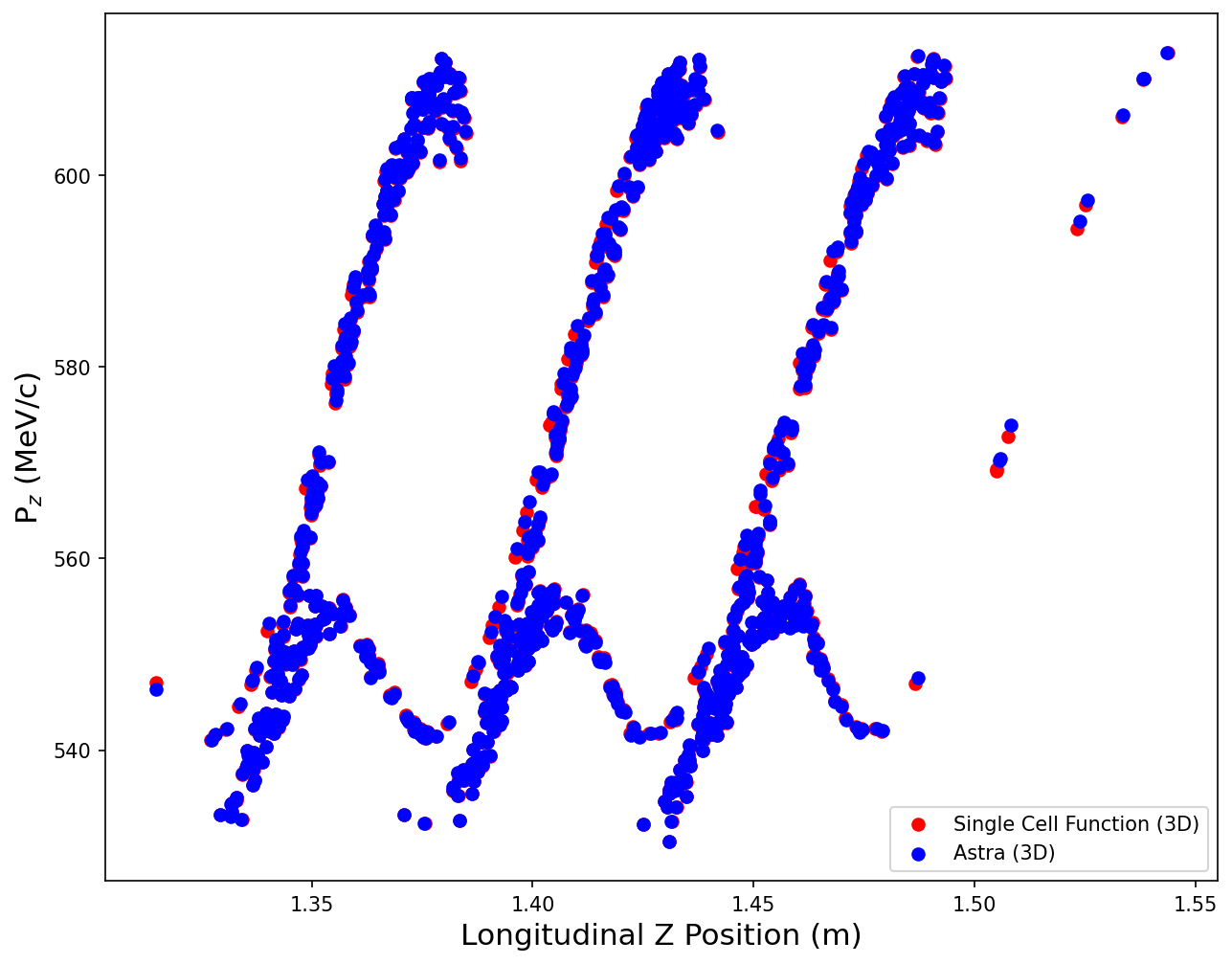}}
     \caption{Transverse and longitudinal phase space of proton beam (initial energy 150 MeV) after traversing 50 SW cells, as calculated by ASTRA and FC2CT.}
     \label{fig:150_protons_sw}
\end{figure}

Figure~\ref{fig:150_protons_sw} displays the output distributions for an initial proton energy of 150 MeV, after 50 SW cells. The two distributions agree well. The largest discrepancy between distributions occurs on the `neck' of the phase space plot. Potential reasons for this are discussed later.

\begin{figure}[ht]
\centering
     \subcaptionbox{}{\includegraphics[width=2.5in]{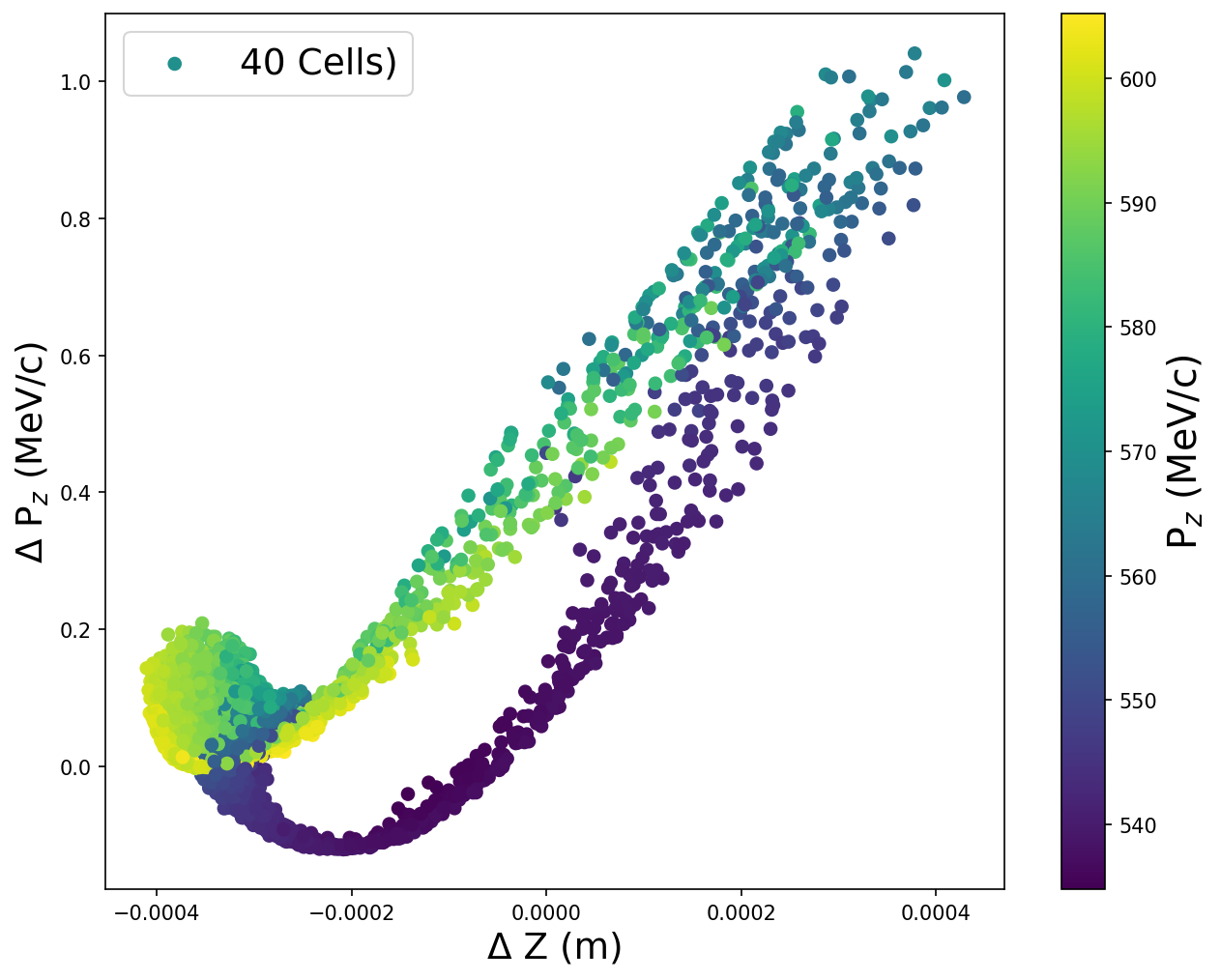}}
     \caption{Displays the difference in calculated $P_z$ as a function of difference in calculated $z$ (defined as $X_{ASTRA} - X_{FC2CT}$), after 40 SW cells. The initial proton energy is 150 MeV.}
     \label{fig:delta_phase_space_sw}
\end{figure}

The difference between phase space components after a 40 cell SW structure, as calculated by ASTRA and FC2CT, are plotted in Fig.~\ref{fig:delta_phase_space_sw}. The plot displays the maximum difference in $P_z$ occurs at maximum (positive) difference in $z$. In this region, FC2CT underestimates both longitudinal particle momentum and position. This result is intuitive, as lower momentum particles will travel shorter distances. The maximum longitudinal momentum error is $~$1 MeV/c, resulting in a maximum percentage error of approximately 0.016~$\%$ in $P_z$. The ASTRA time step was set to 0.001~ns, resulting in approximately 150 iterations per cell.

\begin{figure}[ht]
\centering
     \subcaptionbox{}{\includegraphics[width=2.5in]{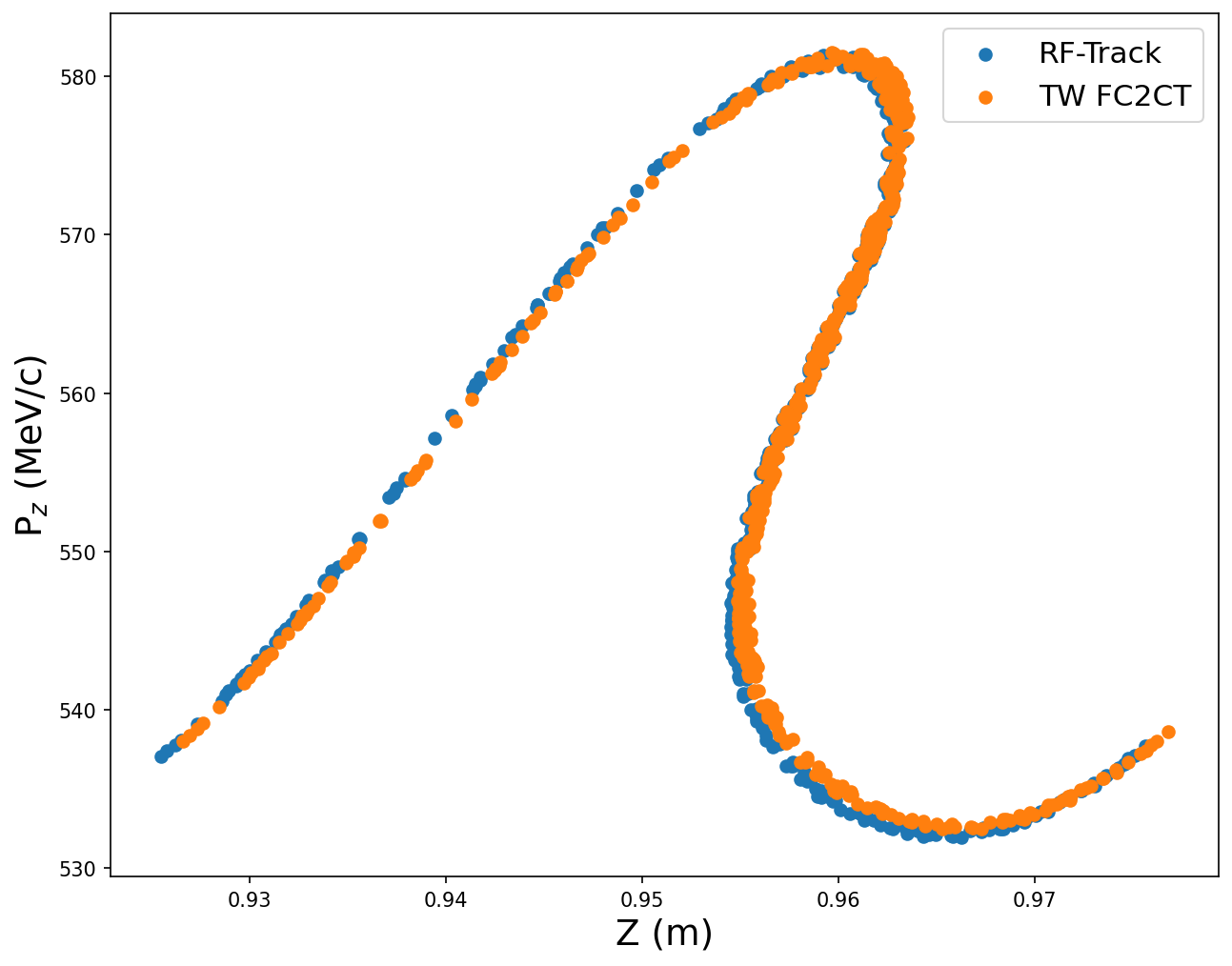}}
     \subcaptionbox{}{\includegraphics[width=2.5in]{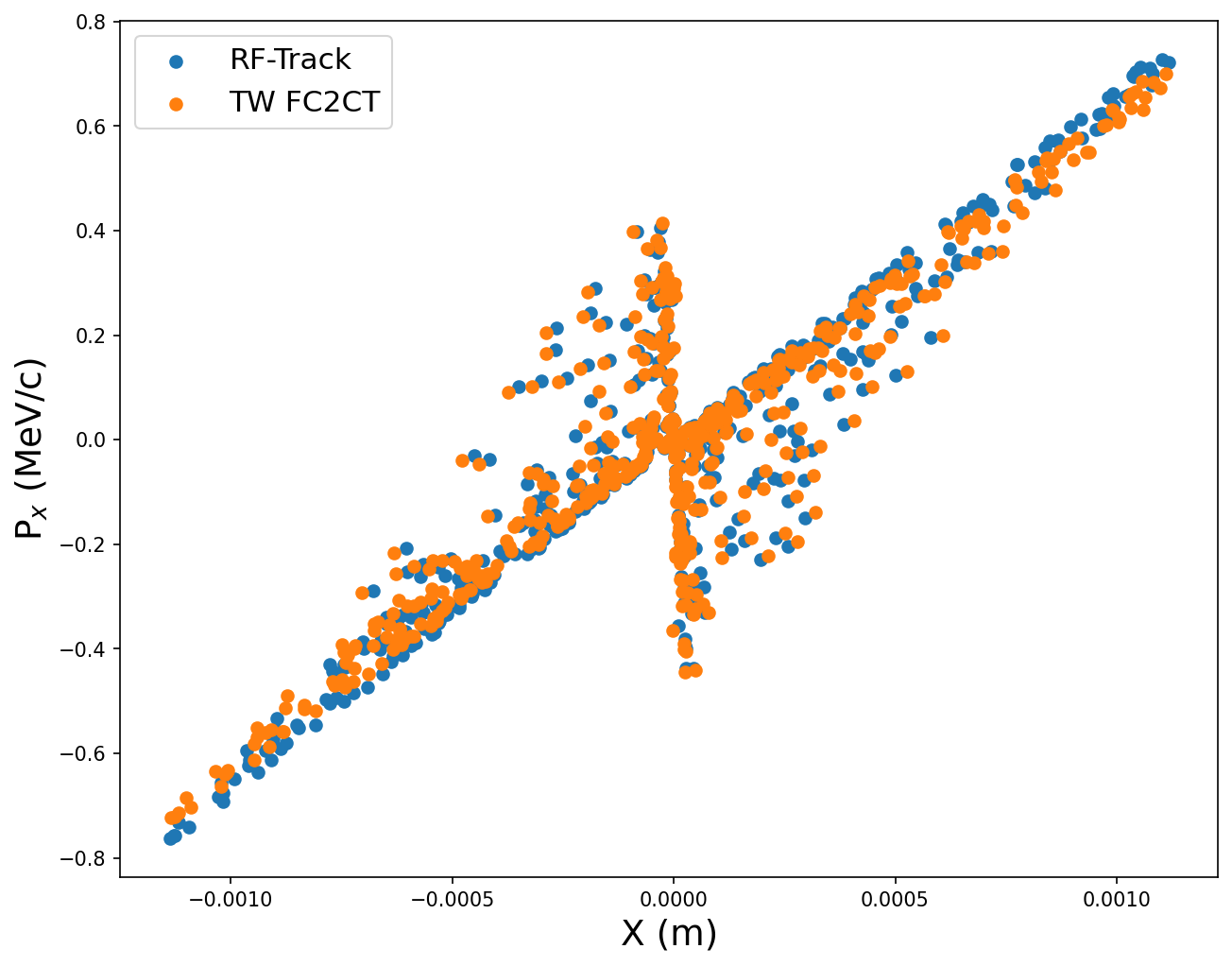}}
     \caption{Transverse and longitudinal phase space of proton beam (initial energy 150 MeV) after traversing 45 TW cells, as calculated by RF-Track and FC2CT.}
     \label{fig:phase_space_tw}
\end{figure}

Particles in this region reside on the `neck' of the phase space plot, see Fig.~\ref{fig:150_protons_sw} (b). These particles begin with gentle acceleration, due to a slightly accelerating initial phase. As the particles begin to phase slip, the phase tends to a zero-crossing phase. In the following cells, particles will experience very small changes in $P_z$, and particles become physically spread out, as bunching occurs. The input distributions of Figures~\ref{fig:37.5_protons} and ~\ref{fig:150_protons_sw} were uniform in rf phase, however particles in the `neck' are spread out, implying thus small changes in initial conditions (rf phase) result in large changes in output conditions. As a result, FC2CT will perform less accurately on particles in this region.

Very small discrepancies in calculated $P_z$ occurs at maximum (negative) error in $z$. In this region, particle position is overestimated by FC2CT. However, the discrepancy in $z$ is very small, with a maximum percentage error of approximately 0.04$\%$. Particles with maximum and minimum $P_z$ are most accurately tracked using FC2CT. This is due to the particles residing at an rf phase that remains far from the zero-crossing phase, where FC2CT performs most poorly.

Figure~\ref{fig:phase_space_tw} displays the longitudinal and transverse phase space as calculated by RF-Track and FC2CT for a TW cavity comprised of 45 cells. The longitudinal electric field component at some time, $t$, is shown in Fig.~\ref{fig:TW_field} (b). The FC2CT method closely agrees with the well trusted tracking code, RF-Track. The output distribution has traversed a 45 cell TW cavity with $4\pi/5$ phase advance per cell. For protons at 150 MeV, the length of the cavity is approximately 1~m. The RF-Track time step was set to 0.003~ns, resulting in approximately 44 iterations per cell. It is shown later that 44 iterations is beyond the limit of convergence.

\begin{figure}[ht]
\centering
     \subcaptionbox{}{\includegraphics[width=2.5in]{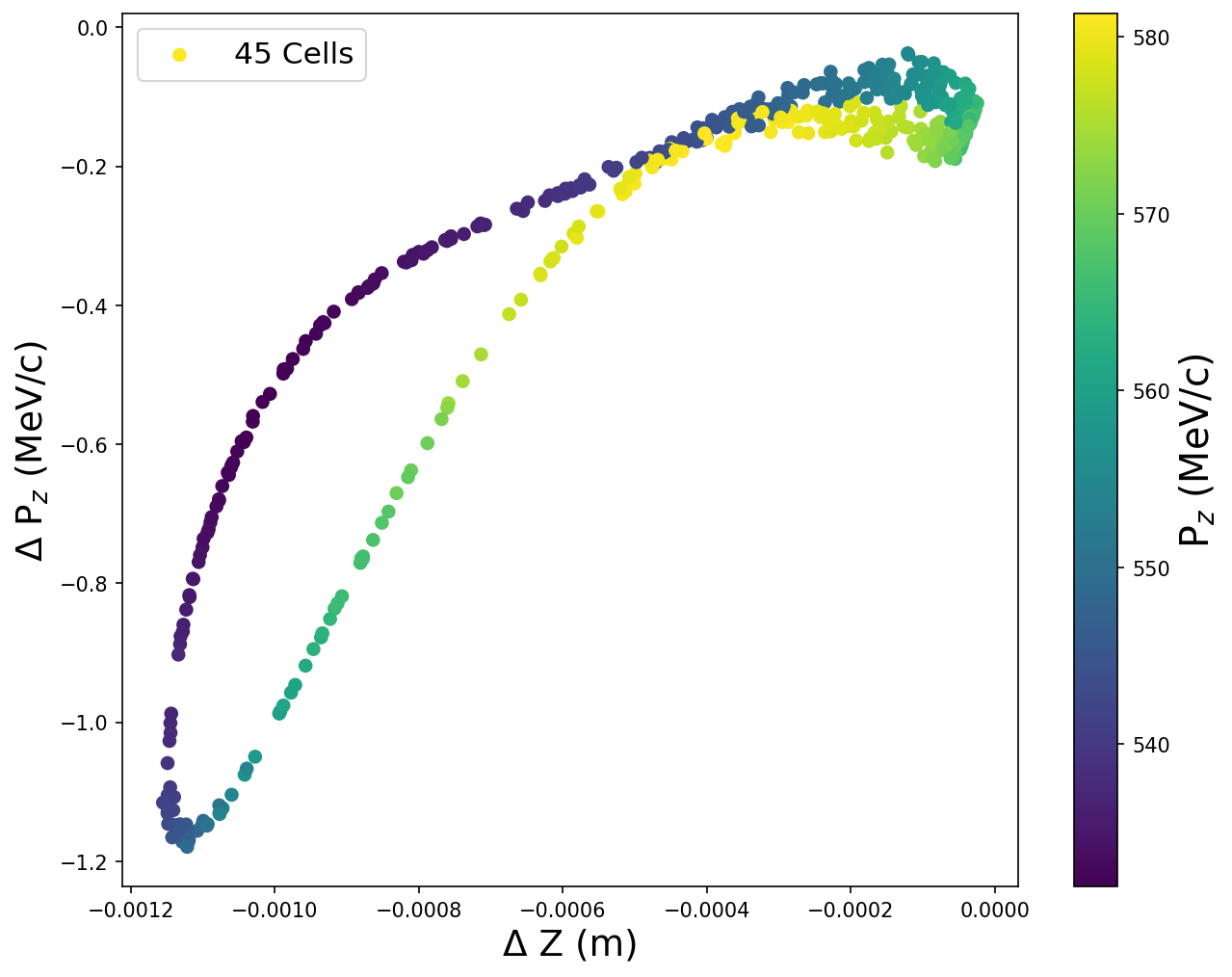}}
     \subcaptionbox{}{\includegraphics[width=2.5in]{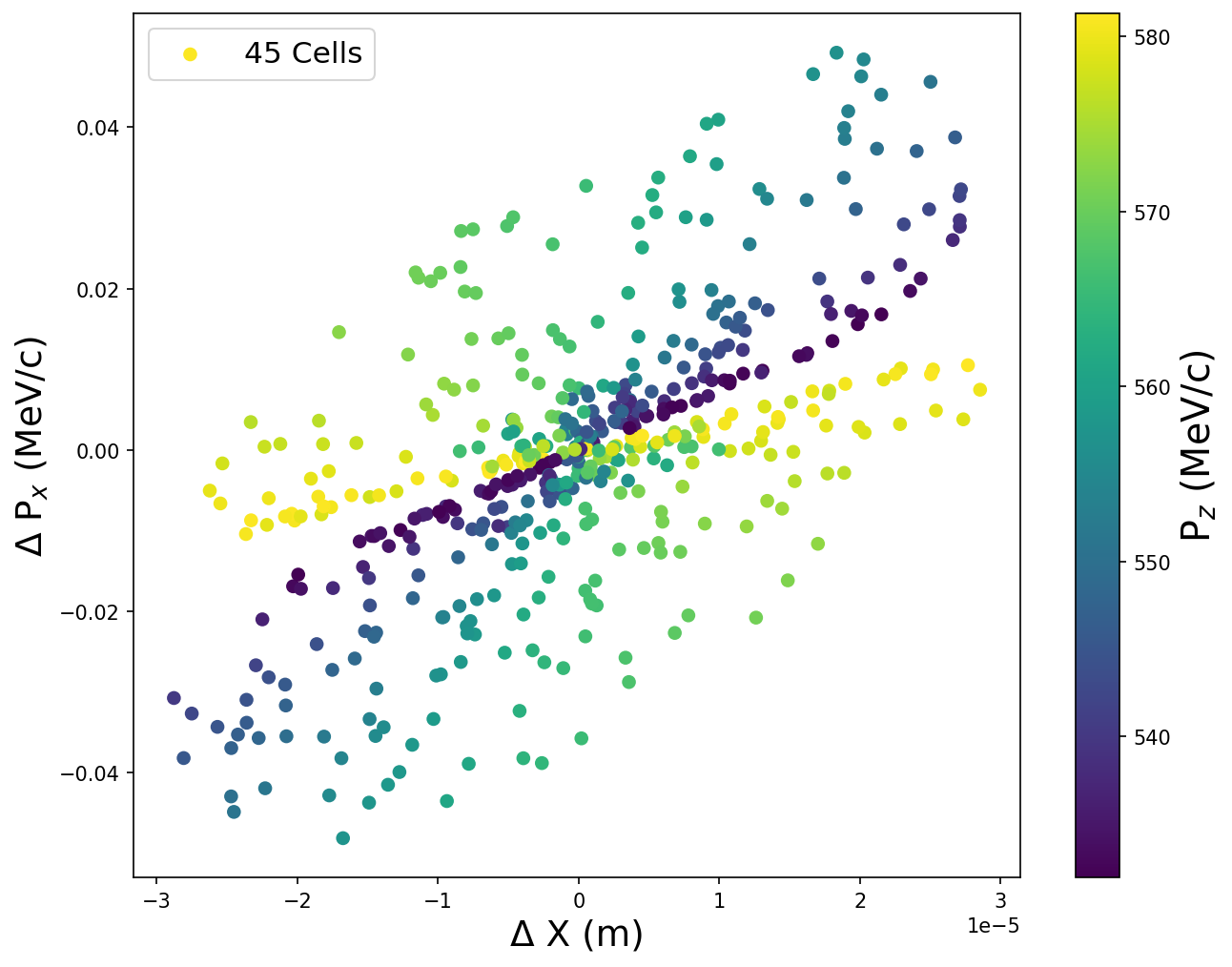}}
     \caption{Sub-figure a/b displays the difference in calculated $P_z/P_x$ as a function of difference in calculated $z/x$ (defined as $X_{ASTRA} - X_{FC2CT}$), after 45 TW cells. The initial proton energy is 150 MeV.}
     \label{fig:delta_phase_space_tw}
\end{figure}

Figure~\ref{fig:delta_phase_space_tw} displays the difference in calculated phase space components between FC2CT and RF-Track. Fig.~\ref{fig:delta_phase_space_tw} (a) shows a similar error relationship to the SW case shown in Fig.~\ref{fig:delta_phase_space_sw}. The majority of particles have small discrepancies in $P_z$, with the phase space of `neck' particles being less well determined. The maximum difference in $P_z$ is approximately 0.02~$\%$, similar to the error in the SW case. The main difference is that FC2CT overestimates both $P_z$ and $z$ for all particles, as $\Delta P_z$ and $\Delta z$ are always negative. Similar to the SW case, maximum $P_z$ particles are well approximated, with small errors in both position and momenta. The maximum position error is $\sim$ 0.0012~m, approximately 6$\%$ of a cell length.

Figure~\ref{fig:delta_phase_space_tw} (b) displays the discrepancy in transverse momentum as a function of transverse position discrepancy. The relationship between errors appears to be a function of $P_z$. For low $P_z$ particles (deep purple), $\Delta P_x$ is a straight line with respect to $\Delta X$. For high $P_z$ particles, the relationship is also linear, albeit with a shallower gradient. For all other particles, $\Delta P_x$ is generally larger, with a maximum discrepancy of approximately 6$\%$. The maximum position error is $\sim$ 1.2 $\%$ of the beam pipe radius.

From results displayed, the FC2CT is a highly efficient and accurate tracking method. The FC2CT method has the ability to vastly reduce the computational expense of large tracking jobs. Whilst FC2CT has been limited above to iterations every rf cell, the method allows iterations over any distance, by changing the bounds of integration in Eqn.~\ref{eq:delta_pz2}.

\subsection{Changes to FC2CT}

FC2CT is based upon the approximation of constant particle $\beta$ over the cell. This approximation is highly effective for multiple energy ranges for both electrons and protons. The accuracy of the approximation is directly proportional to the change in particle $\beta$ over an rf cell, $\Delta \beta$. However, for very low energy protons, $\Delta \beta$ may be small over an rf cell, however $\beta$ may be changing by a factor of two, thus the accuracy of the approximation must be normalised to the particle $\beta$, giving $\frac{\Delta \beta}{\beta}$.

\begin{figure}[ht]
\centering
     \subcaptionbox{}{\includegraphics[width=2.5in]{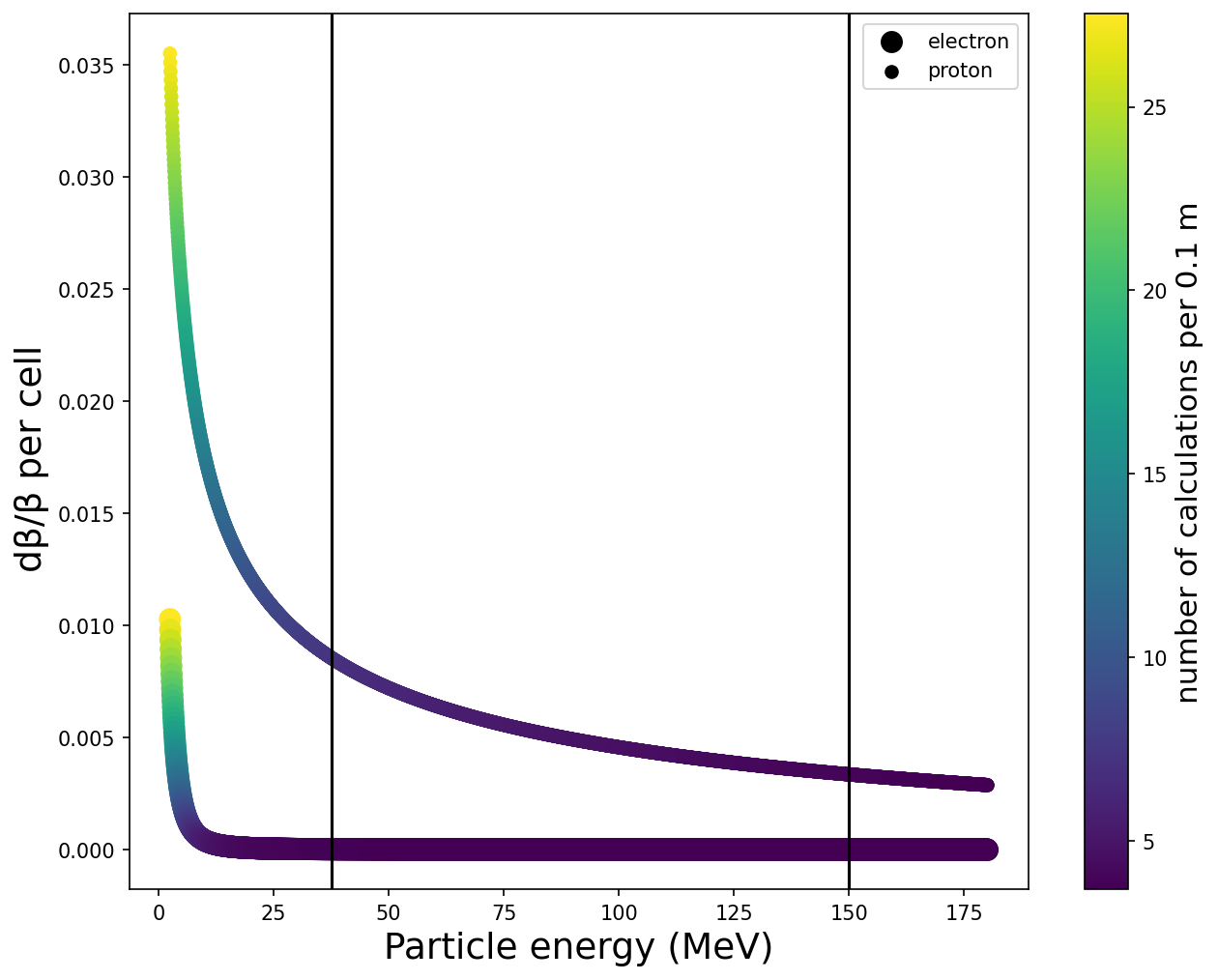}}
     \caption{$\frac{\Delta \beta}{\beta}$ as a function of particle energy. Eye guides are shown at 37.5 and 150 MeV. The starting energy is 2.5 MeV.}
     \label{fig:delta_beta_over_beta}
\end{figure}

Figure~\ref{fig:delta_beta_over_beta} shows the ratio $\frac{\Delta \beta}{\beta}$ for electrons and protons at different energy. The value of $\frac{\Delta \beta}{\beta}$ is calculated by estimating the energy gain over an rf cell of appropriate cell length with a fixed gradient. The quantitative value of $\frac{\Delta \beta}{\beta}$ is thus irrelevant, but the relative value is important for inferring particle energy where the constant $\beta$ approximation is valid.
The eye guides shown at 37.5 and 150~MeV display the region of the phase space plots shown in Figures~\ref{fig:37.5_protons},~\ref{fig:150_protons_sw} and~\ref{fig:phase_space_tw}, where the FC2CT model is valid and accurate. The value of $\frac{\Delta \beta}{\beta}$ is similar for electrons at 2.5~MeV and protons at 37.5~MeV. As previously discussed, CCL are often appropriate for protons with energy beyond $\sim$~30~MeV, as below RFQ and DTL are superior.

\begin{figure}[ht]
\centering
     \subcaptionbox{$\frac{2\pi}{3}$ disk-loaded TW cavity.}{\includegraphics[width=2.5in]{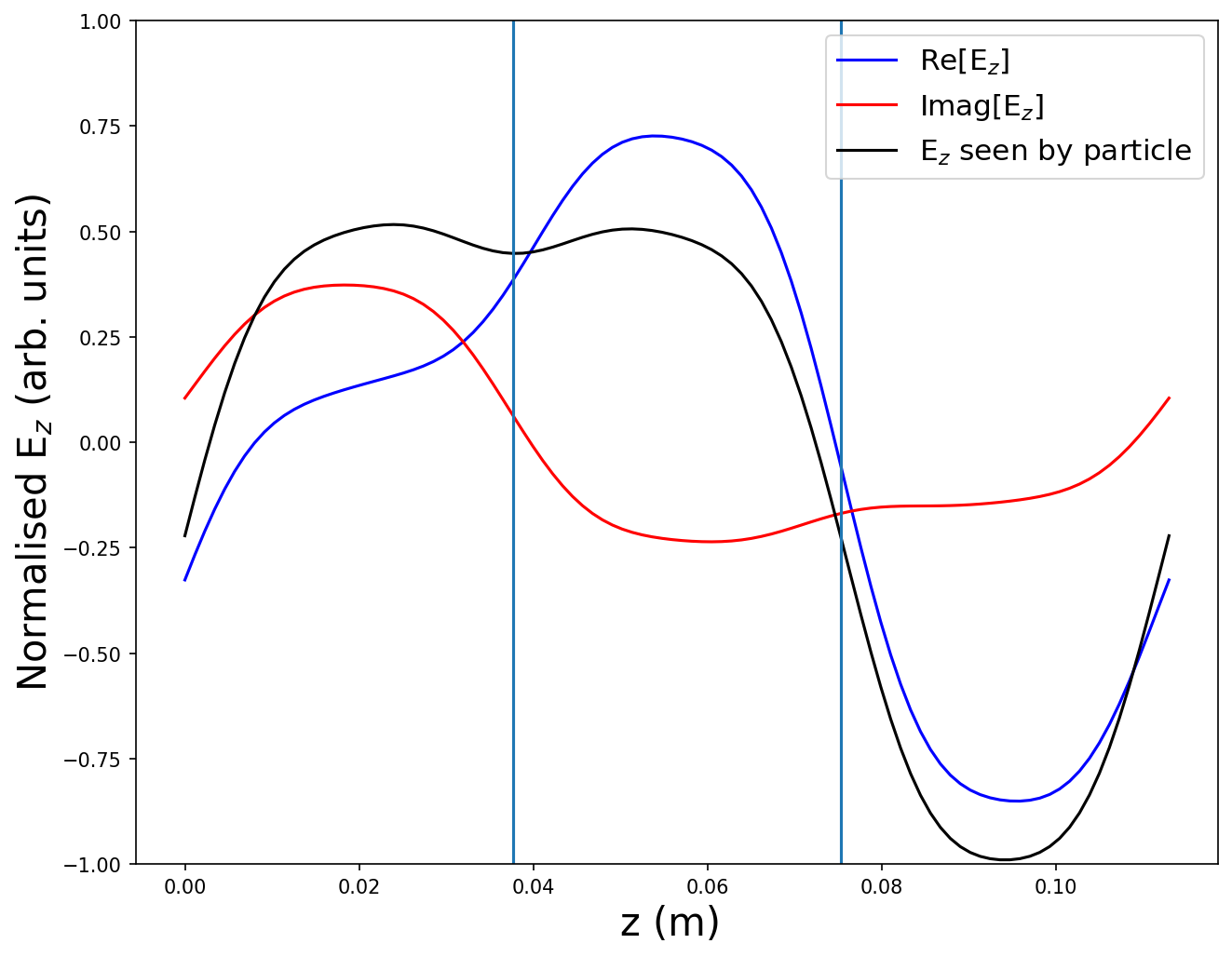}}
     \subcaptionbox{$\frac{4\pi}{5}$ small aperture TW cavity.}{\includegraphics[width=2.5in]{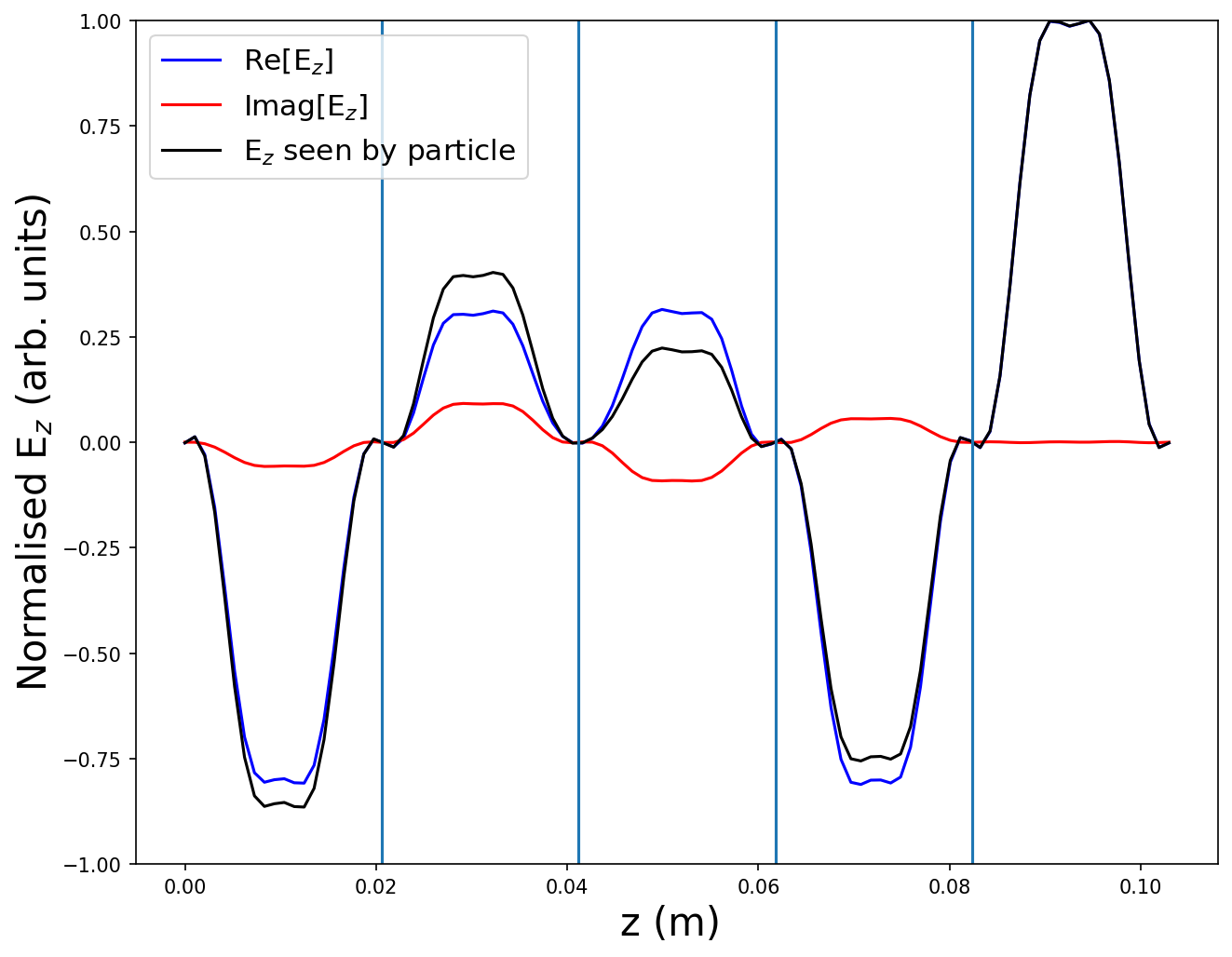}}
     \caption{$E_z$ field for two TW cavity types.}
     \label{fig:TW_field}
\end{figure}

The FC2CT method takes approximately four times longer for tracking through TW structures, as previously described. For certain TW structures, the beam aperture is small, in order to maximise the shunt impedance. For these structures, the cut-off frequency in the beam pipe is very low, and the field is highly evanescent in this region between cells, as is the case for the TW field in Fig.~\ref{fig:TW_field} (b). In these cases, the TW cell can be approximated as a SW cell, utilizing the same method outlined in Section~\ref{sec:sw_gain}. The field has a fixed phase advance per cell and a symmetric on-axis $E_z$ field about the centre of the rf cell, thus is not a true TW. However, the results can be highly accurate, whilst faster by a factor of $\sim$~4.

\begin{figure}[ht]
\centering
     \subcaptionbox{}{\includegraphics[width=2.5in]{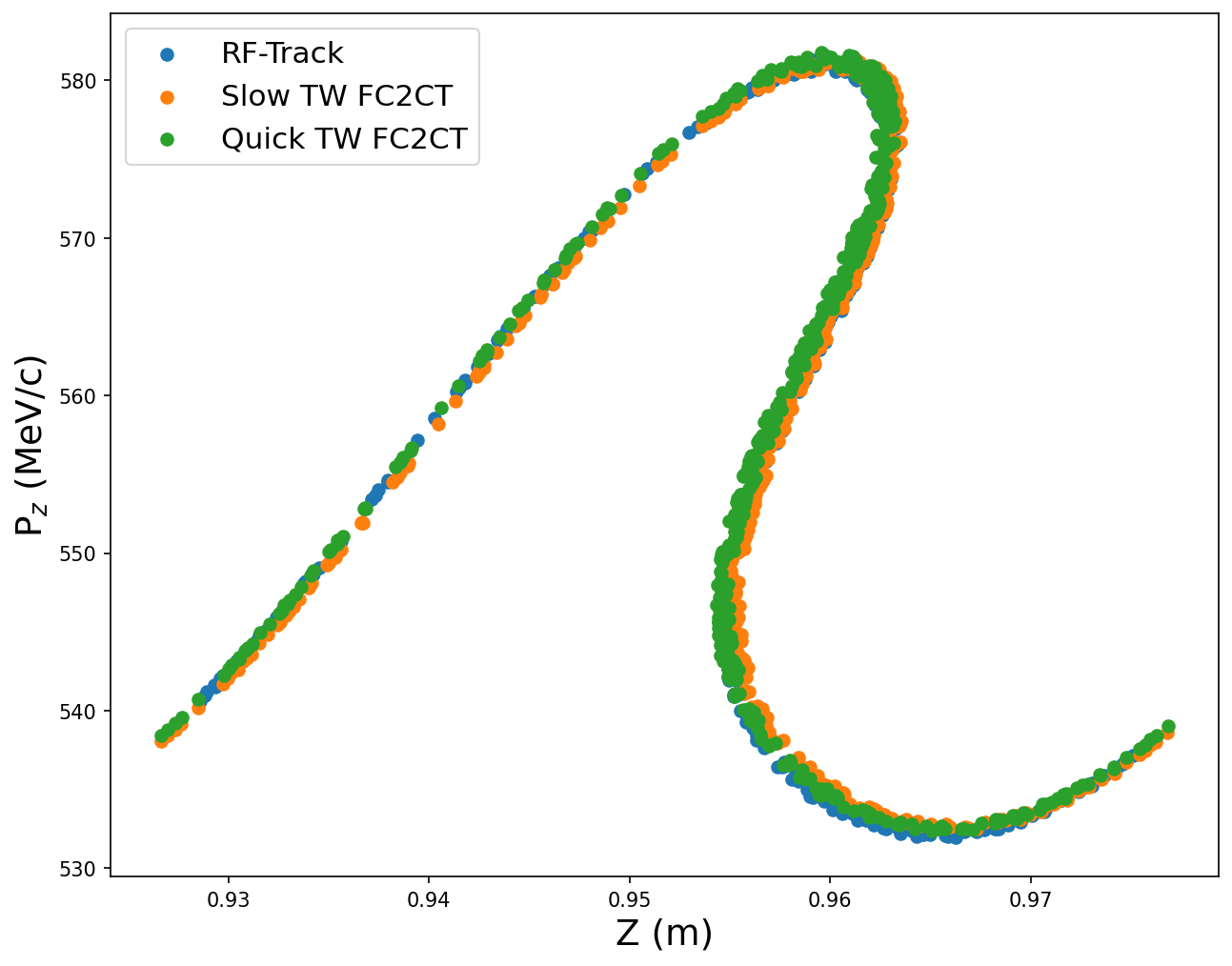}}
     \subcaptionbox{}{\includegraphics[width=2.5in]{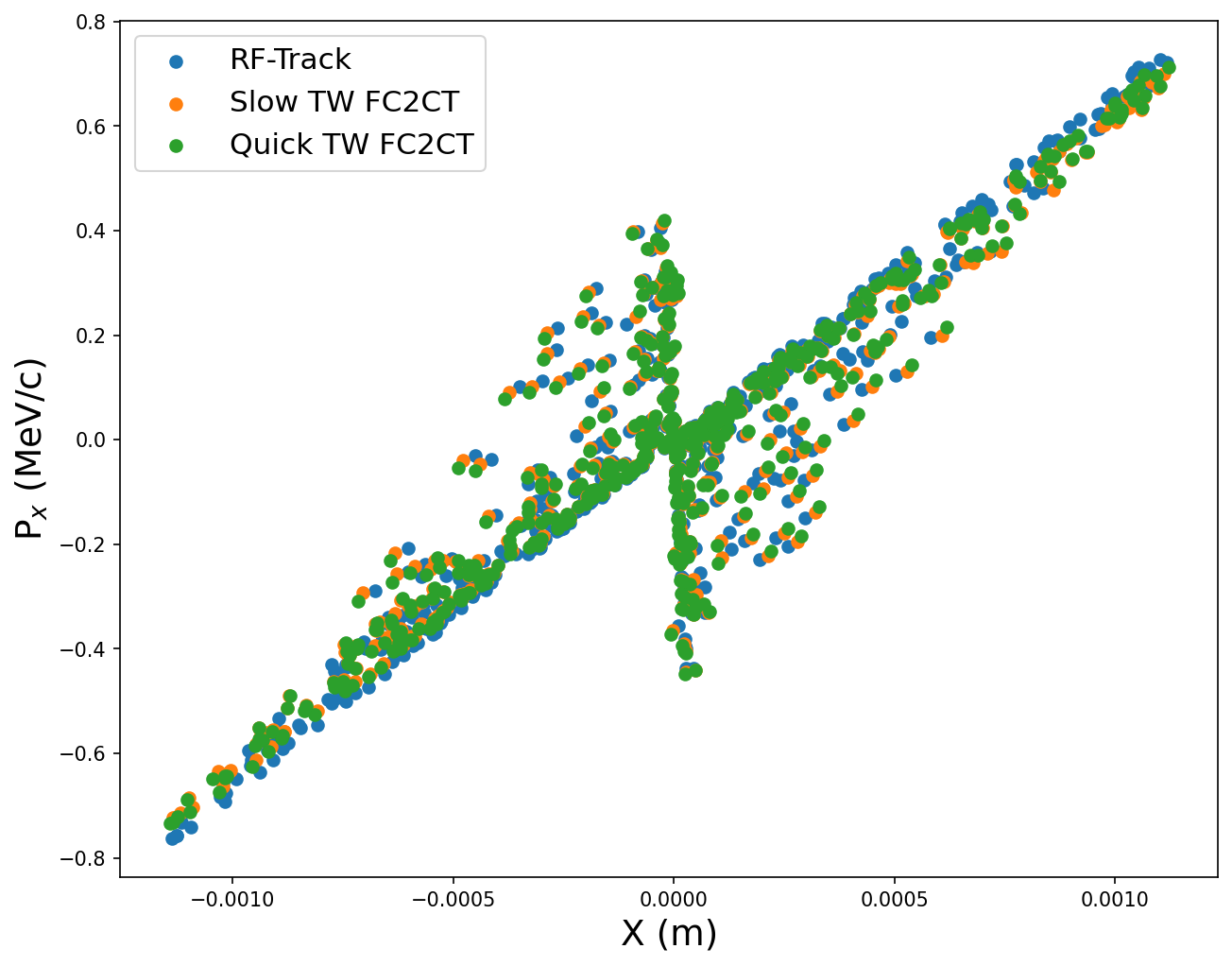}}
     \caption{Transverse and longitudinal phase space of proton beam (initial energy 150 MeV) after traversing 45 TW cells, as calculated by RF-Track and FC2CT as TW (slow) and SW (quick).}
     \label{fig:TW_phase_space_fast_slow}
\end{figure}

Figure~\ref{fig:TW_phase_space_fast_slow} displays the phase space of a proton distribution after traversing a 45 cell TW structure. The true TW FC2CT method (slow) is shown alongside the SW FC2CT method (quick). Whilst the true TW FC2CT agrees with RF-Track more closely, there is not a huge drop in accuracy for the quick method.

Not all TW structures can be treated as SW structures, however. The more asymmetric the $E_z$ field is inside a TW cell, the less accurate the approach becomes. Figure~\ref{fig:TW_field} (a) shows $E_z$ for a disk-loaded electron cavity. For this type of cavity, approximating the field as a symmetric SW is not appropriate, and the true TW FC2CT is required. However, the number of Fourier terms required in the Fourier series description of A(z) and B(z) is far fewer than for the case in Fig.~\ref{fig:TW_field} (b), saving computation time.

Accuracy improvements for the FC2CT model have been explored by 1) running an iteration over an rf cell twice, updating the particle $\beta$ with a more accurate (mean) value for the second iteration or, 2) reducing the integration length of the iteration.

\begin{figure}[ht]
\centering
     \subcaptionbox{}{\includegraphics[width=2.5in]{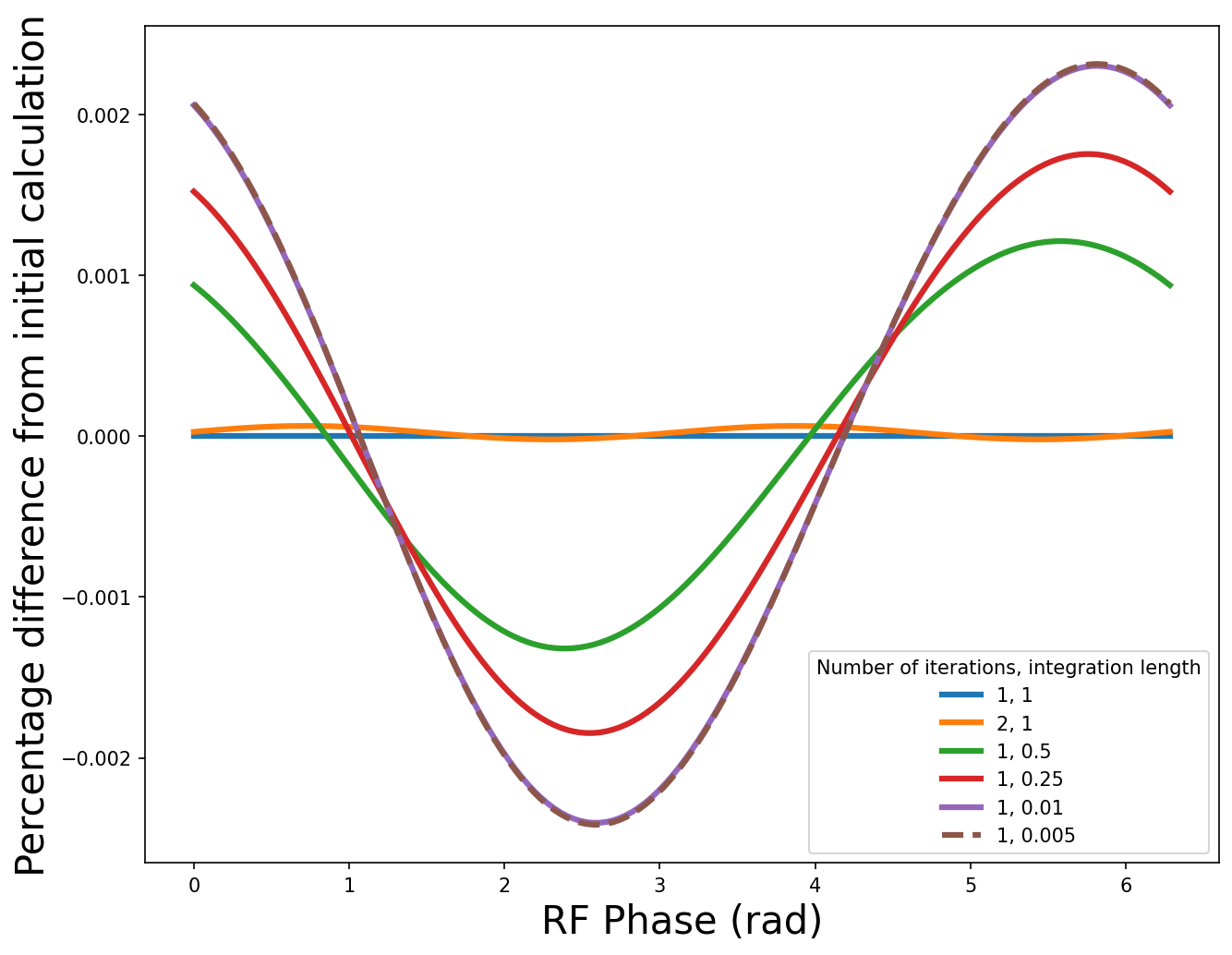}}  
     \caption{The change in initial 1D calculation of $P_z$ for protons at 150 MeV due to 1) updating particle $\beta$ with a mean value, or, 2) increasing the number of iterations per rf cell by decreasing the integration length.}
     \label{fig:FC2CT_improvements}
\end{figure}

Figure~\ref{fig:FC2CT_improvements} displays the change in initial $P_z$ calculation (in a 1D scheme) due to changes in FC2CT parameters, such as number of iterations and integration length, as described in points 1) and 2). The integration length is given as $\frac{L_{iteration}}{L_{cell}}$, so an integration length of 0.5 computes two iterations per cell.

Figure~\ref{fig:FC2CT_improvements} shows that updating the particle $\beta$ with a mean value does not change the value of $P_z$, relative to changing the integration region. The green line shows the change in $P_z$ from running over half an rf cell twice, requiring the same number of calculations as the orange line (two iterations, but updating $\beta$ for the second iteration). A much larger change occurs when reducing the integration step, relative to updating $\beta$.

The change in $P_z$ converges for an integration length of 0.01. Therefore, separating the cell into ten separate calculations is as accurate as the model can be made, beyond which the change is negligible.

\begin{figure}[ht]
\centering
     \subcaptionbox{}{\includegraphics[width=2.5in]{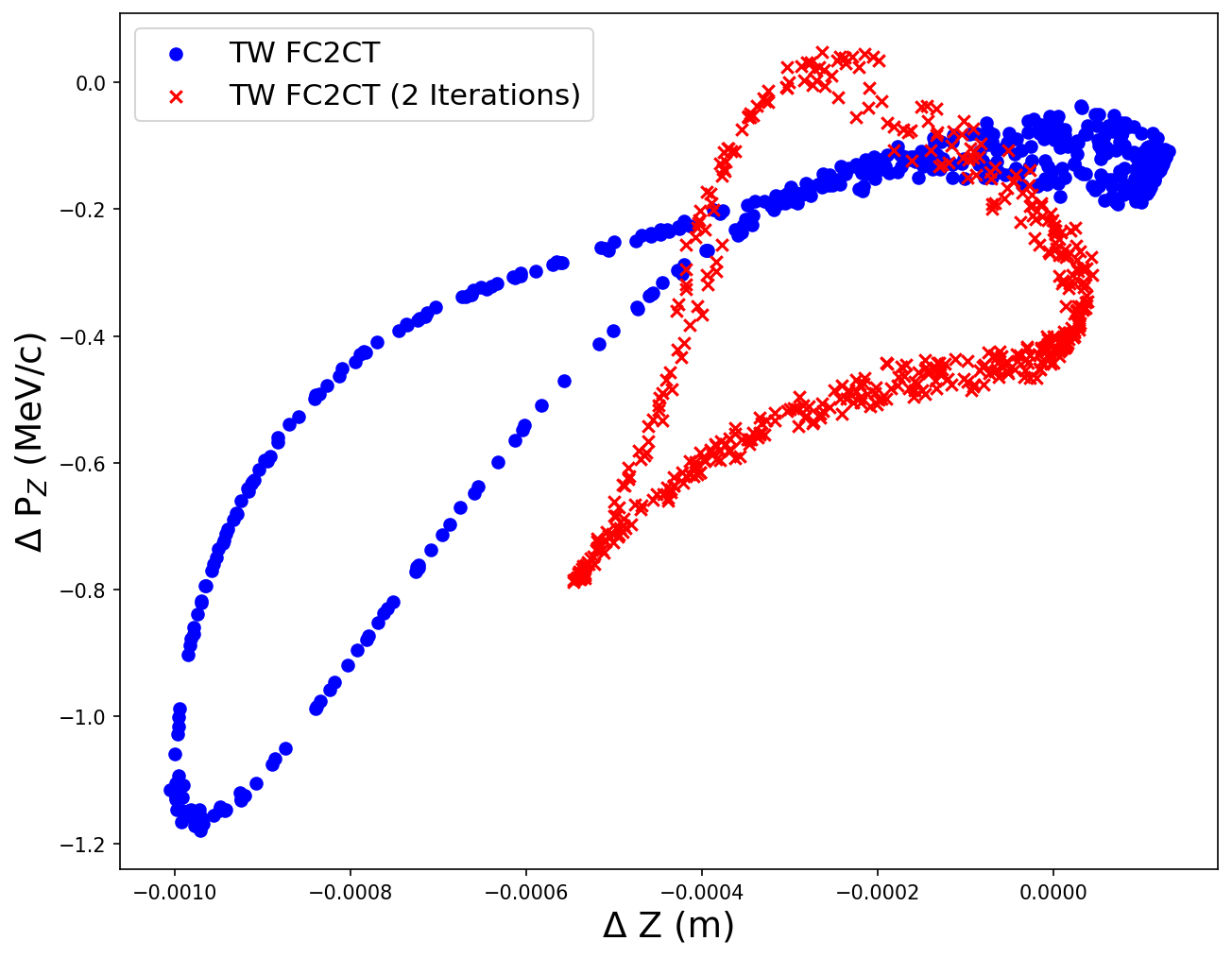}}
     \subcaptionbox{}{\includegraphics[width=2.5in]{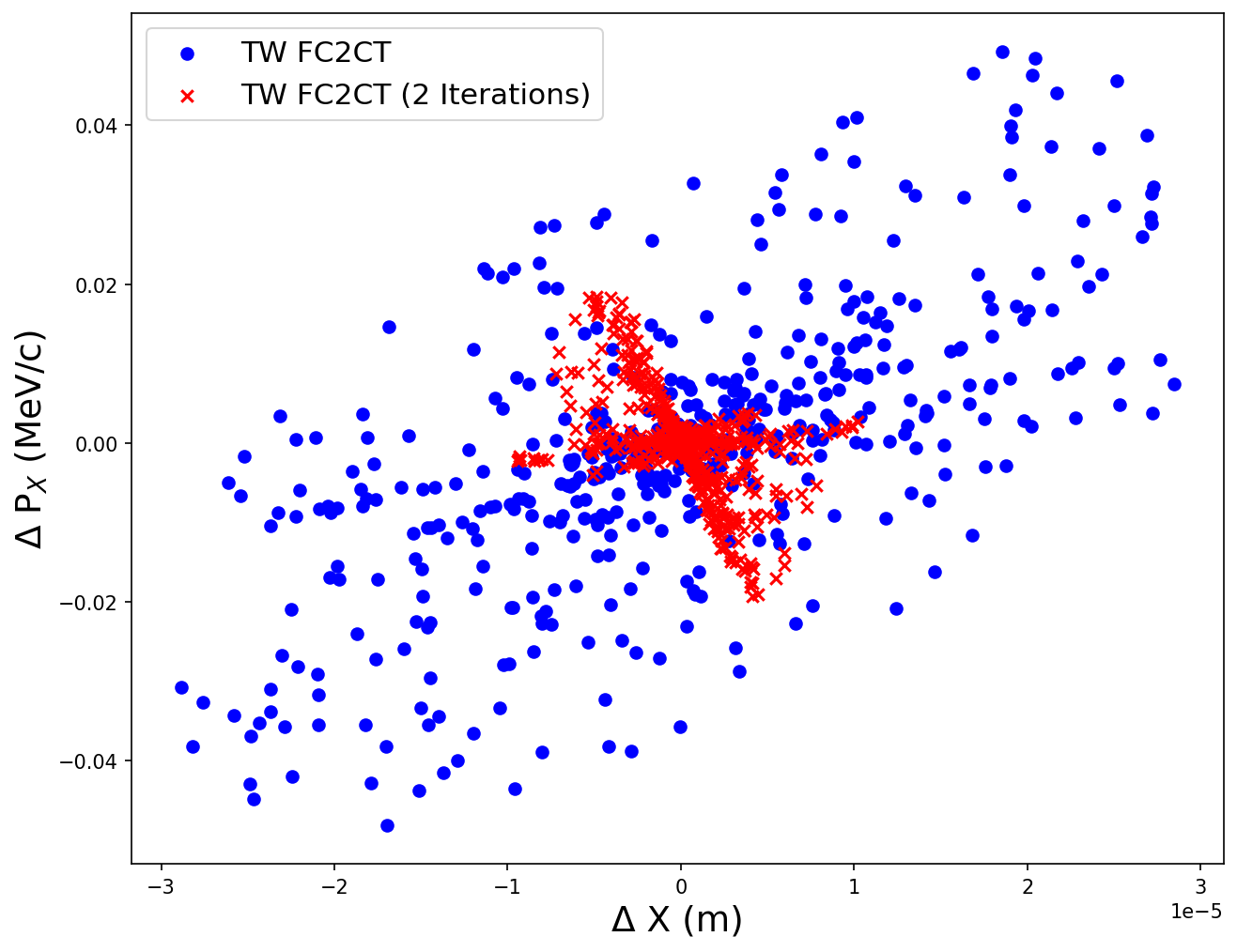}}
     \caption{Error in longitudinal and transverse phase space between RF-Track and FC2CT. Two versions of FC2CT are shown, with one (blue) and two (red) iterations per cell.}
     \label{fig:FC2CT_improvements_2iter}
\end{figure}

Figure~\ref{fig:FC2CT_improvements_2iter} displays the error in phase space between RF-Track and FC2CT, with one and two iterations per cell. As expected, computing two iterations per rf cell produced an output distribution closer to RF-Track.

It is possible to reduce the computation time for FC2CT considerably for both SW and TW structures by using only the principal harmonic.

As FC2CT was designed as fast tracking code, decreasing the integration length is not advised, as the increase in accuracy is minor, with a maximum change in $P_z$ of $\approx$~0.002 $\%$, requiring the computation time to increase by a factor of ten. In addition to reducing the integration length, computing time could be further reduced by iterating over multiple cells in one calculation. In general, the integration length must be chosen to reflect the particle rest mass and energy, in addition to the required accuracy. Another time-saving option for FC2CT is to use only the first harmonic in the Fourier Series expansion. This is possible as higher order spatial harmonics cancel exactly for relativistic particles, and are very small for non-relativistic particles.

\section{Conclusion and Next Steps}
\label{section:conclusion}

In this paper, the FC2CT method for fast particle tracking was displayed, with its performance compared relative to two well trusted tracking codes, ASTRA and RF-Track. FC2CT demonstrated strong agreement with the codes for both SW and TW cavities. Additional developments for FC2CT were discussed, such as altering the integration length, and treating certain TW cavities as SW cavities.

The SW FC2CT method works very effectively, as the number of computations is low, relative to TW FC2CT. Whilst TW FC2CT was highly accurate as a particle tracker, the computation time is approximately four times longer than the SW case. Currently, the effect of fringe fields at the entrance/exit of an rf cavity is not accounted for. Further work can incorporate the radial impulse due to these fringe fields to better approximate a real life situation.

In order to quantitatively measure the reduction in computation time, an FC2CT mode must be constructed in a compiled coding language. In this study, FC2CT was written in an interpretive language, the script is not written with minimising computation time as a major objective.

\begin{acknowledgments}
The studies presented have been funded through the Cockcroft Core Grant by STFC Grants No. ST/P002056/1 and ST/V001612/1.
\end{acknowledgments}


\bibliographystyle{plain} 
\bibliography{main} 
\end{document}